# Electric-Field-Induced Tautomerism in Metal-Free Benziporphyrins Enables Aromaticity-Controlled Conductance Switching

Yenni Ortiz-Acero[1], Arnau Cortés-Llamas[1], Jordi Ribas-Arino[1], Stefan T. Bromley*[1,2]

[1]*Departament de Ciència de Materials i Química Física & Institut de Química Teòrica i Computacional (IQTCUB), Universitat de Barcelona, c/Martí i Franquès 1-11, 08028 Barcelona, Spain*

[2]*Institució Catalana de Recerca i Estudis Avançats (ICREA), Passeig Lluís Companys 23, 08010 Barcelona, Spain*

* Corresponding author: s.bromley@ub.edu

## Abstract

Metal-free porphyrins can switch between hydrogen-bonded tautomers, potentially enabling reversible control in molecular electronics. However, electric field gating of porphyrin tautomerism, which is critical for device integration, has not been fully realized. We propose metal-free benziporphyrins (MFBPs), in which one pyrrole ring is replaced with a phenol group, as a new platform for tautomer-based molecular electronics. This approach introduces asymmetry, which allows for three distinct tautomers, each possessing a characteristic aromatic or antiaromatic electronic structure. Density functional theory and quantum transport calculations show that: i) experimentally realisable electric fields can selectively stabilize each tautomer, and ii) each tautomer exhibits a characteristic conductance profile. The strong switching capability of MFBPs is demonstrated by ON/OFF ratios exceeding 500 at low bias. Fused MFBPs further expand functionality by providing multiple tautomeric states for multistate molecular registers and enabling wire-like architectures with enhanced conductance. These results establish MFBPs as versatile building blocks for electric-field-responsive molecular devices and open new research opportunities for molecular-scale logic and memory.

## Introduction

Porphyrins have attracted considerable attention in molecular device research due to their chemical stability, extended π-conjugation, and ease of functionalization [1, 2]. Their ability to form stable metal complexes via coordination of the four inner nitrogen atoms is widely exploited to tune electronic and optical properties for molecular devices [3, 4] and functional materials [5]. In metal-free porphyrins (MFPs), two hydrogen atoms occupy the inner nitrogen sites. These hydrogens undergo rapid tautomerism, effectively transferring as protons while preserving the neutral, fully conjugated π-electron system of the macrocycle. This dynamic behavior has minimal impact on the delocalized π-electron magnetism, making MFPs promising candidates for molecular spintronic applications [6, 7]. However, tautomerism can locally modulate electronic properties, enabling MFPs to serve as platforms for molecular switching via controlled proton transfer [8, 9, 10, 11]. Such tautomeric switching is particularly appealing for device applications because it (i) generates well-defined, distinct molecular states, (ii) is reversible, and (iii) occurs without significant structural perturbation of the porphyrin macrocycle.

For practical device switching, the tautomers in MFPs must be accessible via external stimuli. This control was first demonstrated by varying the current through single MFPs adsorbed on a surface [8], and conductance-based tautomerism has also been observed in single-molecule junction experiments [11]. These studies, together with theoretical work [9], show that different tautomers exhibit distinct electronic transport properties. For device applications, control via electric-field (E-field) gating is highly desirable. Using a scanning tunneling microscope tip, it has been shown that the barrier for tautomerism can be lowered by applied E-fields [10]. However, selective gated control of MFP tautomers by E-fields has not yet been achieved. One potential limitation is the high symmetry of MFPs: many tautomers are rotationally equivalent and have low or zero dipole moments, reducing coupling to external fields.

To explore more asymmetric tautomerism, we consider metal-free benziporphyrins (MFBPs), in which one five-membered pyrrole ring is replaced by a six-membered π-conjugated ring [12, 13]. This substitution introduces a slight out-of-plane distortion and, when suitably functionalized, provides an additional site for proton transfer. As a result, MFBPs can host a broader range of structurally and energetically distinct tautomers. We show that these increased degrees of freedom create a versatile platform for highly tunable, E-field-responsive molecular devices.

Herein, we investigate an experimentally synthesized metal-free benziporphyrin (MFBP), 22-hydroxybenziporphyrin [14], which incorporates a phenol fragment and exhibits tautomerism. Unlike typical MFPs, this MFBP is structurally asymmetric, enabling access to three symmetrically inequivalent tautomers, which couple directly to changes in aromaticity. The two most stable tautomers, T1 and T2, rapidly interconvert in solution, while a slightly higher-energy tautomer, T3 (see Figure 1), likely serves as an intermediate in this dynamic interconversion [14, 15]. Although the chemical differences among these tautomers are subtle, their electronic structures differ markedly: T1 and T3 are antiaromatic around the macrocycle, whereas T2 exhibits local aromaticity around the phenol fragment.

Porphyrinoids have long been studied as systems in which aromaticity can be tuned via structural modification or redox chemistry [16]. In such systems, topology- or redox-induced switching between aromatic and antiaromatic states can strongly influence molecular conductance [17, 18]. In the MFBP studied here, electron delocalization and aromaticity are intrinsically controlled by tautomerism, offering a novel mechanism for modulating electronic structure. We theoretically explore how externally applied E-fields can selectively stabilize specific tautomers, and how these tautomer-dependent states produce pronounced changes in electronic transport properties.

We demonstrate, for the first time, that experimentally realizable, suitably oriented E-fields can directly gate tautomerism in MFBPs. This gating can selectively stabilize each of the three tautomers thereby controlling the MFBP aromaticity. This E-field–driven switching is independent of any bias current, distinguishing it from previously reported conductance-controlled tautomerism. We then investigate single-molecule junctions for the three tautomers, as well as junctions formed by two fused MFBPs with all six possible tautomeric combinations. In all cases, we observe one or two stepwise jumps in current as the bias voltage increases. The magnitude and voltage dependence of these jumps are specific to each junction, providing distinct, measurable transport signatures that directly reflect the underlying tautomeric and aromatic states. Notably, in single MFBP junctions, the predicted ON/OFF ratio at low bias can reach up to 500, reflecting a strong switching response arising from the interchange between aromatic and antiaromatic tautomers.

Moreover, by modifying the chemical linkers between the MFBPs and the junction contacts, this discrete multistate behaviour can be transformed into a more gradual current–voltage response, characteristic of a molecular wire, accompanied by enhanced coherent transport. Taken together, these results highlight the versatility of MFBPs as E-field–responsive units, capable of functioning both as multistate molecular switches and as tunable molecular wires. Overall, MFBPs provide an enriched range of electronic and transport properties, offering new opportunities to expand the functionality of porphyrinoid-based molecular devices.

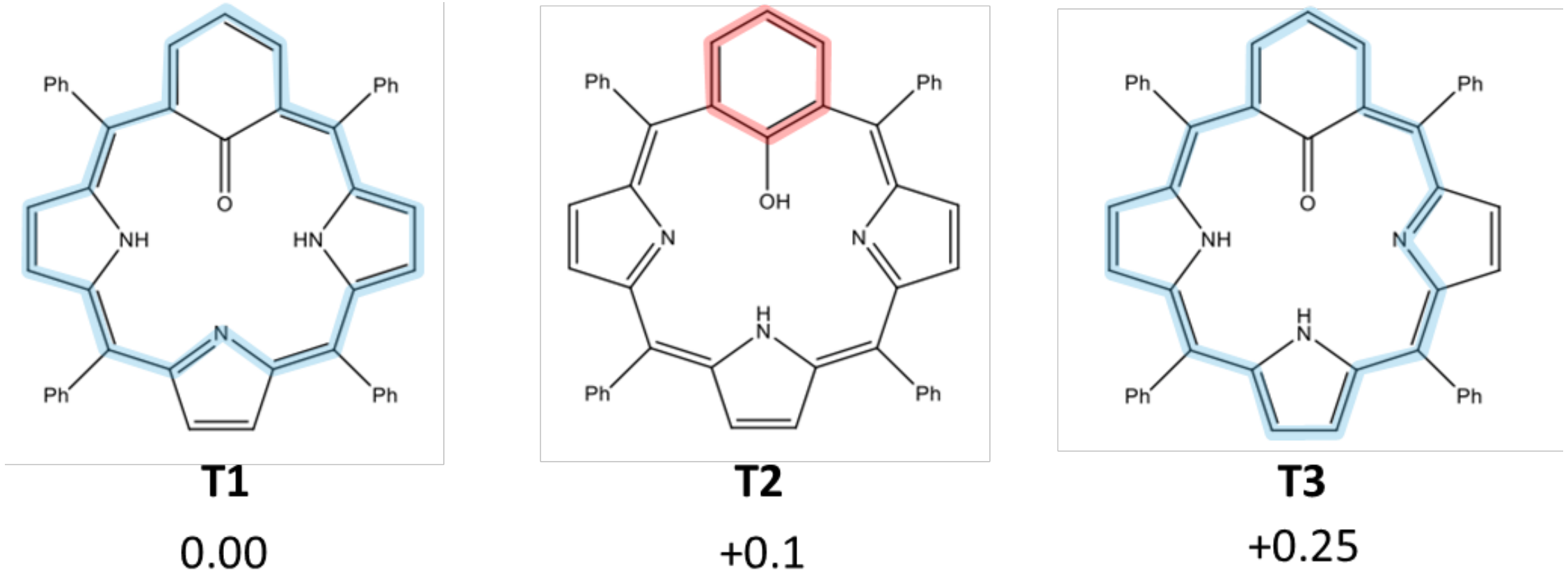

**Figure 1.** Chemical structures of the three considered MFBP tautomers: **T1**, **T2** and **T3**. Relative calculated total energies for zero applied E-field are given in eV with respect to **T1**. The antiaromatic 20 π-electron pathways around the macrocycles are highlighted in blue in **T1** and **T3**. The local 6 π-electron aromatic pathway around the phenol ring is highlighted in red in **T2**.

## Methodology

### *Density functional theory calculations*

Relaxed structures and energies of all freestanding molecules (i.e. without gold electrodes) were obtained via density functional theory (DFT) based optimisations using the PBE0 hybrid exchange-correlation functional [19], and a 6-311g(d,p) basis set, as implemented in the Gaussian09 [20] code. Dispersion interactions were also included using Grimme's D3 scheme [21,22]. The three tautomers were then modelled with respect to their response to E-fields and their electronic transport properties. Applied E-field were included via the "Field" keyword to define the direction and strength of an applied uniform E-field. In such calculations, a new term accounting for the potential due to the E-field is added to the external potential of the Kohn-Sham Hamiltonian. The Kohn-Sham equations are then solved taking into account the effective potential and thus the relaxation of the electronic density is self-consistently evaluated with influence of the applied E-field. Uniform E-fields were thus applied to all tautomers in two in-plane directions (i.e. with respect to the plane defined by the three N atoms of each tautomer) with field strengths ranging from zero to 0.9 V/Å. To assess the influence of the applied E-field, we evaluated the fractional populations of each tautomer ($P(n)$) for all field strengths and both field directions

$$P(n) = \frac{1}{Z}\, e^{\frac{-E_n}{k_B T}} \qquad (1)$$

where $E_n$ is the relative total energy of each tautomer $n$, $T$ is the temperature (taken to be room temperature, 298,15 K), $Z$ is the total partition function for all three tautomers, and $k_B$ is the Boltzmann constant.

***Quantum electron transport calculations***

Quantum electron transport calculations employed model junctions in which the MFBPs were connected between two gold clusters via sulfur atoms. We employed gold clusters composed of 19 atoms in a face-centered cubic structure cut from the bulk crystal structure. DFT calculations using the B3LYP exchange-correlation functional [23] together with the LANL2DZ basis set were employed to treat this junction region. LANL2DZ combines effective core potentials (ECPs) with double-zeta quality basis functions for the valence electrons of the Au atoms, with an explicit all-electron basis set for the lighter elements of the MFBPs. This choice allows for a reliable representation of the electronic structure while minimizing artifacts (e.g. ghost transmission) that can arise in transport calculations [24]. The explicitly represented Au contacts were coupled to virtual semi-infinite leads, to allow for the electron transport through the system to be calculated by means of the nonequilibrium Green's function method (NEGF) [25, 26, 27] as implemented in the ARTAIOS software package [28]. In the NEGF calculations the Green's function of the scattering region (i.e. between the electrodes) $G(E)$ is evaluated following equation 2.

$$G(E) = (E\,S - H - \Sigma_S - \Sigma_D)^{-1} \qquad (2)$$

where $E$ represents the electron energy, $H$ denotes the effective single-particle Kohn-Sham Hamiltonian, and $S$ is the overlap matrix obtained from the DFT optimization. The self-energies $\Sigma_{S/D}$ account for the influence of the source and drain reservoirs, which are coupled to the system via the gold clusters to facilitate electron injection and extraction. These reservoirs are modeled using the wideband approximation [29]. With the Green's function, the transmission function can be calculated as

$$T(E) = 4Tr\left[Im(\Sigma_S)G Im(\Sigma_D)G^{\dagger}\right] \qquad (3)$$

as well as the local transmissions between atoms $i$ and $j$

$$T_{ij}(E\,) = \mathrm{Im}\,(H_{ij}^{*}\,G_{ij}^{n}) \qquad (4)$$

where the correlation function is given by

$$G^n = 2G\ \mathrm{Im}(\Sigma_S)G^{\dagger} \qquad (5)$$

The current $I$ is determined by integrating the energy-dependent transmission function $T$. The range of integration depends on the equilibrium Fermi energy $E_F$ of the electrodes and the applied bias voltage ($V$).

$$I(V) = \frac{2e}{h}\int_{E_F-\frac{eV}{2}}^{E_F+\frac{eV}{2}} dE\ T(E,V) \qquad (6)$$

where $e$ is the unit electronic charge and $h$ is Planck´s constant.

The local transmission between pairs of atoms $i$ and $j$ is calculated from the imaginary part of the product of the Hamiltonian matrix element $H_{ij}$ and the corresponding element of the correlation function. Summing these atomic-pair contributions reproduces the global transmission, while retaining spatial resolution of current pathways.

## Results and Discussion

### *E-field gated molecular switching*

In figure 2 we show how the fractional population of each tautomer varies with respect to applied E-field strength for two indicated in-plane E-field directions and for field strengths from zero to 0.9 V/Å. We note that such E-field strengths are achievable with devices based on dual ionic liquid gating that can generate up to 0.4 V/Å,[30] or via STM tips which can generate E-fields up to 2.0 V/Å.[31]

For low E-field strengths for both considered field directions T1 is the dominant tautomer, as expected from it relatively high zero field stability (see Fig. 1). For a -90° applied E-field (Fig. 2a), the main competition is between T1 and a minor but slowly increasing fraction of T2. This situation persists until a field strength of approximately 0.4 V/Å, above which the fractional population of T2 starts to more rapidly increase, with a corresponding decrease in the fraction of T1. At an E-field strength of ~0.69 V/Å the fractional weight of the two tautomers interchanges. This T1→T2 switch is further reinforced with increasing E-field strength.

For the 135° E-field direction (see Fig. 2b) we observe a small gradual decrease of the dominance of T1 with increasing field strength up to 0.4 V/Å due a small increase in the fraction of T2. For E-fields above 0.4 V/Å the prevalence of T2 falls away while T3 starts to strongly compete with T1. At an E-field strength of ~0.63 V/Å there is a crossover between the fractional populations of T1 and T3 such that T3 becomes the dominant tautomer. As with the -90° case, an increasing E-field strength leads to a reinforcement of T1→T3 switching.

For all tautomers the applied E-field causes only very minor structural distortions to the MFBP skeleton indicating that the E-field-induced proton transfer would be an relatively efficient process. We also find that the observed E-field induced switching can be qualitatively rationalised by computing the relative energies of the induced dipoles of each tautomer in each applied field scenario (see SI).

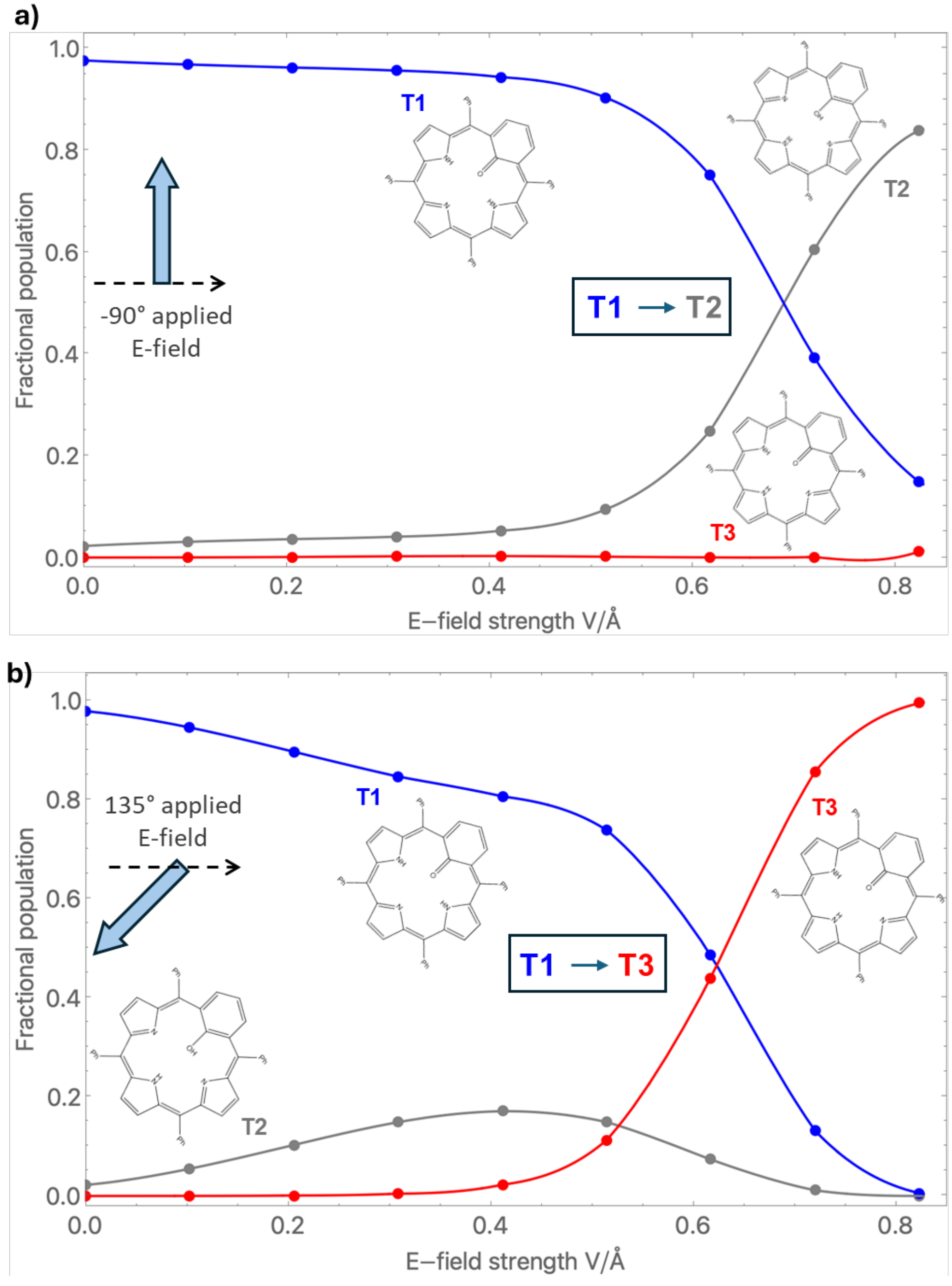

**Figure 2.** Fractional populations for each tautomer at 298.15 K with respect to the applied E-field strength for two field directions: a) -90°, and b) 135°. Note that the field directions are given relative to the connection axis used in the molecular junctions (see dashed arrows). The MFBP inset figures of each tautomer are also rotated so that they are aligned with the same connection axis.

### ***Quantum transport through MFBP junctions***

We first consider transport through junctions based on single tautomers ($\mathbf{J}_{\mathbf{T}n}$) connected to the gold electrodes via two opposing pendant aryl rings (see example in Fig. 3a). The calculated current versus voltage curves (Fig. 3b) in each case show a single step-like increase in current at a specific bias voltage which then plateaus afterwards. Of the three junctions, $\mathbf{J}_{\mathbf{T2}}$ clearly has the highest current carrying capacity for the considered bias voltages. The voltages at which these junction-specific current increases occur correspond to the maximum transmission energy, which is found to correspond to the energy of the highest occupied molecular orbital (HOMO) of the system in each case (see Fig. 3c). This is consistent with a transport regime dominated by coherent, molecule-centered transport. The main difference between the HOMOs of these tautomers is the involvement of the six-membered ring. In T1 and T3, the *p* orbitals of the ring's carbon atoms (particularly those bonded to the methine bridges and the carbon atom attached to oxygen) make substantial contributions to the HOMO. In contrast, T2 exhibits only a minor contribution from the carbon atom bonded to oxygen, leading to negligible overlap between the macrocycle's π-system and that of the six-membered ring. The limited orbital participation of the six-membered ring distinguishes T2 from T1 and T3.

Local current analysis (see Fig. 3c) also reveals that $\mathbf{J}_{\mathbf{T2}}$ exhibits the most direct conduction pathway, with minimal branching of electron flow. This indicates that charge transport in $\mathbf{J}_{\mathbf{T2}}$ proceeds predominantly along a single conjugated route connecting the electrodes, reducing scattering and destructive interference. In contrast, the more bifurcated current pathways in $\mathbf{J}_{\mathbf{T1}}$ and $\mathbf{J}_{\mathbf{T3}}$ display lower overall conductance, reflecting the impact of competing routes and partial current cancellation within the molecular framework. These tendencies are likely linked to the local aromaticity of T2 of the six-membered ring versus the extended macrocycle antiaromaticity of **T1** and **T3** (see Fig. 1). The stabilizing effect of aromaticity is often associated with lower single-molecule conductance, thus tending to favour antiaromaticity for improved junction performance [32, 33]. Here, indeed, we observe a detrimental effect of the local aromaticity of the phenol ring fragment in **T2** (see Fig. 1b), which tends to inhibit transport through it. However, this also leads the transport to take a more direct (non-branching) path through the MFBP molecule. Conversely, when the six-membered ring is participates in an antiaromatic macrocycle (i.e. **T1** and **T3 –** see Fig. 1a, c), an increase of branching currents through the ring are permitted leading to an overall decrease in direct source-to-drain transport.

In figure 3b we also indicate the tautomeric switching that could be induced by suitably applied E-fields (see Fig. 2). For device applications the E-field induced switching between tautomers should ideally be associated with a corresponding large change in current at a fixed bias voltage. The ratio of the ON (i.e. high) current to the OFF (i.e. low) current measures how effectively a device can switch. For bias voltages in the range 0 - 0.1 V, the current through $\mathbf{J}_{\mathbf{T1}}$ remains relatively low (<1 nA). Conversely, in the same bias voltage range the current in $\mathbf{J}_{\mathbf{T2}}$ rapidly increases to over 3.5 μA. Using an applied E-field to induce **T1**(OFF) ↔ **T2**(ON) switching (see Fig. 3a) would lead to a junction with an ON/OFF ratio around 500 throughout this bias voltage range. For progressively larger bias voltages this ON/OFF ratio decreases but remains above 100 up to a bias voltage of 0.4 V. These absolute currents are expected to be larger than typical experimental single-molecule measurements due to the idealized contact coupling, absence of inelastic scattering, and coherent transport assumed in the NEGF calculations. The ON/OFF ratios, rather than the absolute currents, provide a more robust measure of the intrinsic switching contrast of the MFBP junctions. We note that the finite bias ON/OFF ratio associated with the T2/T1 tautomerism is over order of magnitude higher than that calculated [9] or measured [11] for a similar tautomerism in symmetric MFPs.

For the case of the **T2** ↔ **T3** tautomerism, both tautomers have similar *I vs V* characteristics for much of the bias voltages sampled. The currents appreciably diverge for an approximate bias voltage range of 0.3 - 0.6 V, where switching is thus possible. Here, we observe a maximum ON/OFF ratio of ~38 at a bias voltage of ~0.4 V. At this bias voltage, switching to **T2** is also possible with an ON/OFF ratio of ~100. In principle, this would permit three state switching **T2** ↔ **T1**↔ **T3** with a bi-directional gating field.

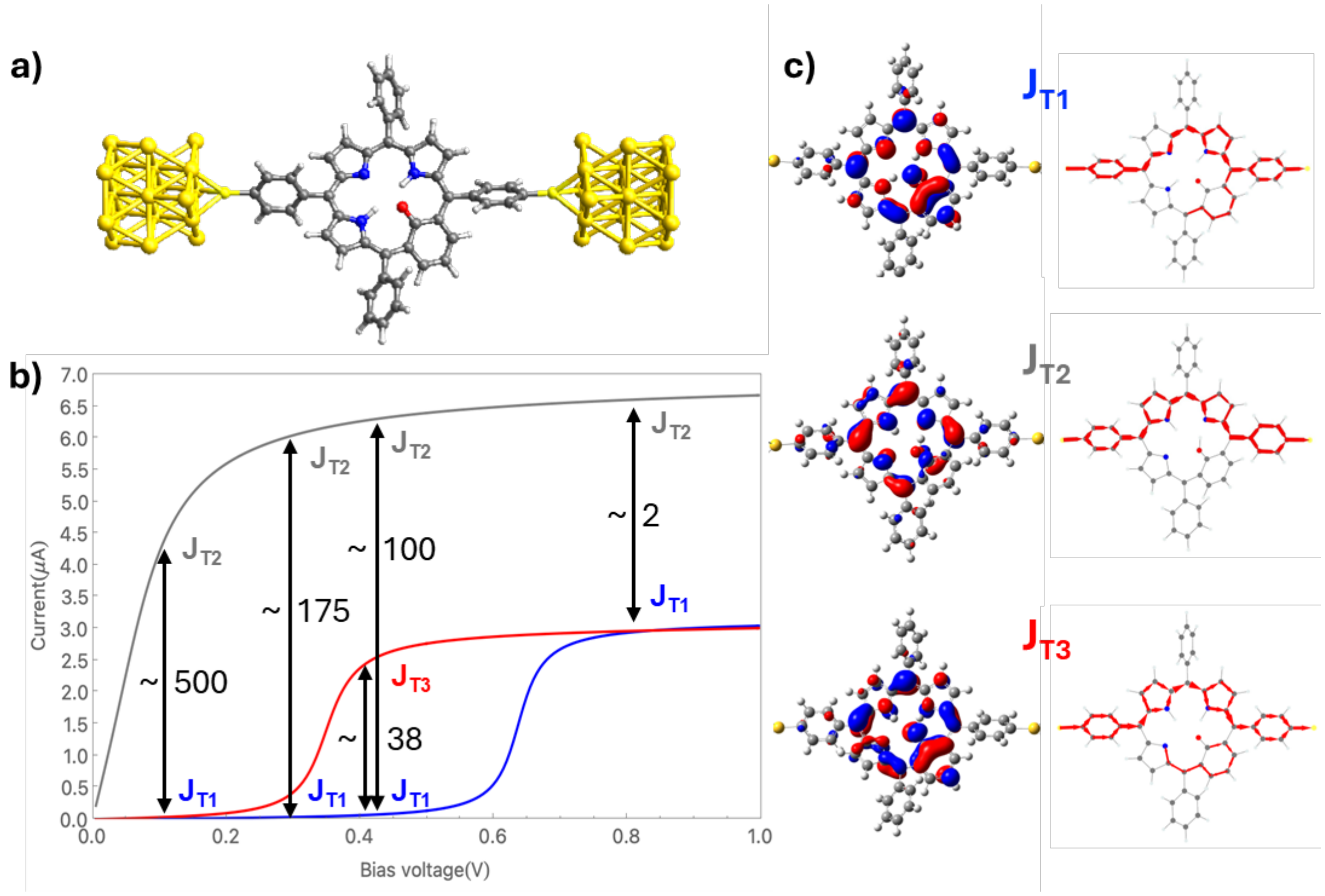

**Figure 3.** a) Structure of molecular junction model with the T1 MFBP tautomer linked to two gold electrodes via sulphur atoms ($\mathbf{J}_{T1}$). Atom colour key: C – dark grey, O – red, N - blue, H - light grey, S - light yellow, Au - dark yellow. b) Calculated current with respect to bias voltage for three single MFBP junctions for junctions: $\mathbf{J}_{T1}$ (blue line), $\mathbf{J}_{T2}$ (grey line) and $\mathbf{J}_{T3}$ (red line). Reversible arrows indicate possible E-field induced tautomeric population shifts following figure 2. Numbers denote approximate ON/OFF current ratios at specific bias voltages. c) HOMO orbitals associated with the maximal transmission in each junction (left) and the corresponding local transport pathways (right).

To further assess the potential of MFBPs as multistate molecular switches, we calculated the transport characteristics of junctions formed from all six possible combinations of two fused MFBP tautomers (denoted $\mathbf{J}_{Tn\text{-}Tm}$). The MFBPs are fused via two pyrrole rings in each tautomer, following the triply linked connectivity commonly employed to construct highly conjugated porphyrin-based molecular wires [5]. Such edge-fused porphyrin architectures, often referred to as *porphyrin tapes*, have been extensively investigated for metal-containing porphyrins, while hybrid tapes incorporating MFPs have also been experimentally realized [34].

Because the six-membered benzenoid ring breaks the symmetry of the MFBP macrocycle, fusion through pyrrole rings requires a relative rotation of the two MFBP units. In all fused junctions considered here, one MFBP is rotated by 90° with respect to the other (see the $\mathbf{J_{T1\text{-}T1}}$ junction in Fig. 4a); the corresponding structural models for the remaining junctions are provided in the Supporting Information.

Similar to the single-tautomer junctions, the fused systems generally exhibit a single stepwise increase in current at a characteristic bias voltage within the range studied, with the exception of the $\mathbf{J_{T1\text{-}T1}}$ junction, which displays a more complex two-step current–voltage response (Fig. 4b). Junctions composed of identical tautomers yield the three highest maximum currents; however, in all cases the fused junctions exhibit significantly reduced current compared to their single-tautomer counterparts. Importantly, all six fused junctions display stepwise current increases at distinct bias voltages and reach different maximum current levels, providing six clearly distinguishable transport signatures. To date, experiments have achieved four-state tautomer-based conductance-based switching (i.e. not gate induced) at high bias voltages and low temperatures for single MFP junctions [7].

These results underscore the potential of MFBP-based junctions as multistate molecular electronic elements. Unlike conventional two-state ON/OFF molecular switches, these systems offer multiple stable, addressable configurations, enabling more complex switching schemes, increased information density, and enhanced functionality for molecular-scale logic and memory applications.

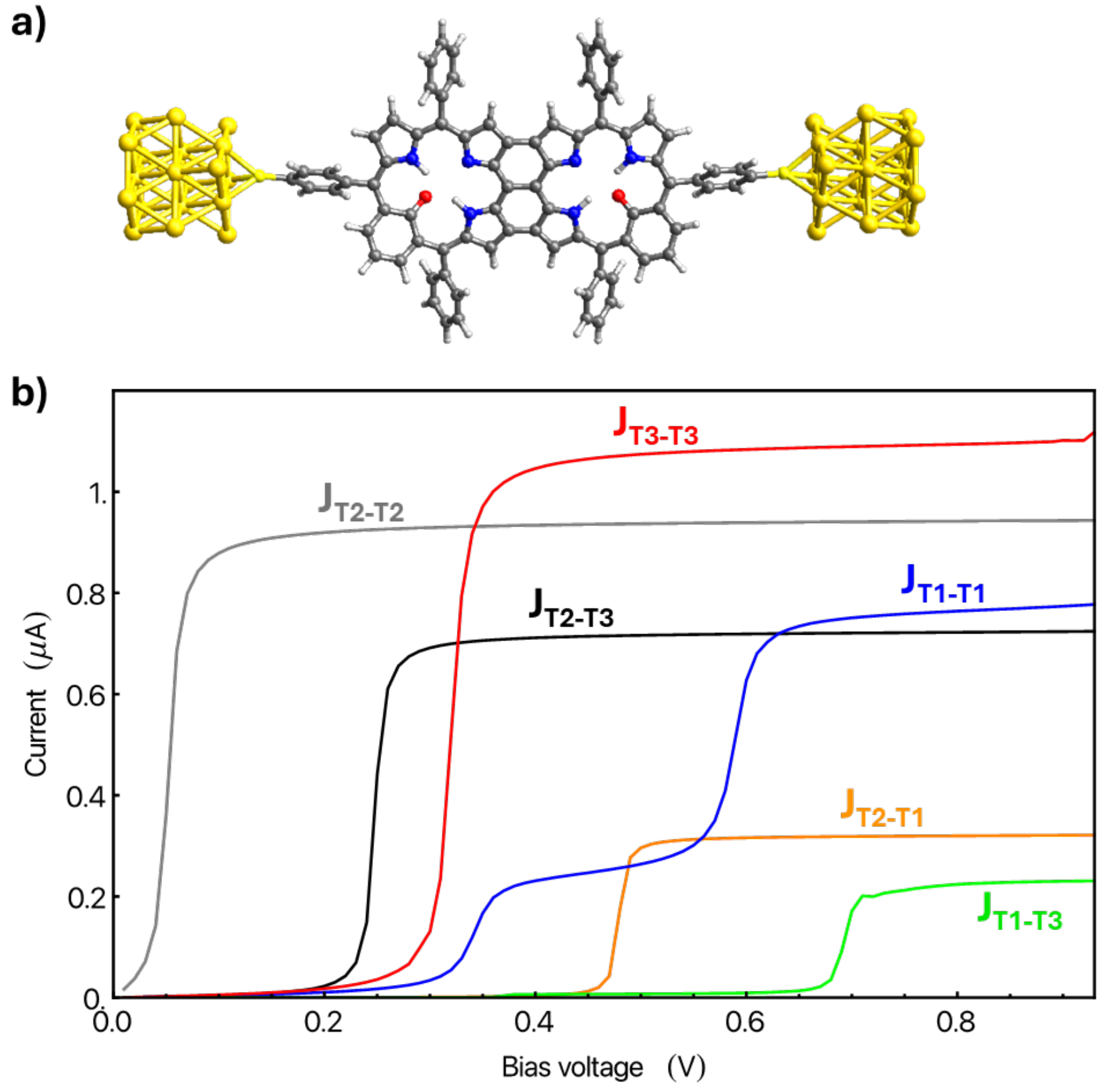

**Figure 4.** a) Structure of molecular junction model with the two fused **T1** MFBP tautomers linked to two gold electrodes via sulphur atoms ($\mathbf{J_{T1-T1}}$). Atom colour key follows that used in figure 3. b) Calculated current with respect to bias voltage for six junctions comprised of fused combinations of **T1**, **T2**, **T3** tautomers.

### *MFBP-based molecular wires*

Unlike switchable molecular junctions, which exhibit stepwise I–V characteristics due to discrete changes in conductance, molecular wire-type junctions typically display relatively high currents that vary smoothly with applied bias, reflecting continuous electron transport through a conjugated pathway. It is well known that the nature of the molecule-electrode connection in porphyrin junctions can significantly affect the observed transport characteristics [35]. Here, we examine these effects in three different variants of $\mathbf{J_{T1}}$ (see Fig. 5), referred to as $\mathbf{J_{T1a}}$ , $\mathbf{J_{T1b}}$ and $\mathbf{J_{T1c}}$. Case $\mathbf{J_{T1c}}$ is the junction shown in Fig. 3a. In case $\mathbf{J_{T1b}}$, we replace the aryl rings linking the molecule with the electrodes with acetylenic linkers. In case $\mathbf{J_{T1a}}$, we further remove two pendant aryl rings. See Fig. 5a for the structure of each junction variant.

For the case $\mathbf{J}_{T1c}$ we see a narrow transmission feature, where the peak at ~0.3 eV corresponds to the single stepped increase of current at ~0.6 V shown in Fig. 3 (see equation 4). This result is in line with the known effect of aryl rings to selectively tune the degree of conjugated π-orbital overlap via their twist angle when used as linkers in junctions [36] and materials [37]. Conversely, for the acetylenic-linked $\mathbf{J}_{T1a}$ and $\mathbf{J}_{T1b}$, we see a significant broadening and shifting of the transmission function peaks. Acetylenic linkers provide rigid, linear, and highly conjugated connections between molecular backbones and electrodes, leading to strong molecule–electrode coupling and significant orbital hybridization. As a result, transmission resonances are substantially broadened compared with junctions containing more weakly coupled twisted aryl linkers. We note that the local transport pathways in each case (see Fig. 5c) is relatively unaffected by the nature of the connection of the MFBP to the electrodes.

As a minimal models of MFBP molecular wires we consider variants of $\mathbf{J}_{T1\text{-}T1}$, $\mathbf{J}_{T2\text{-}T2}$ and $\mathbf{J}_{T3\text{-}T3}$ junctions in which we employ acetylenic linkers to the gold electrodes. In figure 6 we compare the current versus voltage characteristics of these junctions with those of their single molecule equivalents. We note that in all cases, the current through acetylenic-linked single-molecule junctions is at least one order of magnitude higher than that in the aryl ring linked single-molecule junctions (see Fig. 3b). In all cases the edge-fused MFBP wires show a more complex pattern of increasing current with respect to bias voltage as compared to the single molecular junctions. While the current of in the single molecule systems tends to start levelling off at around 0.5 V, the current in the wire-like systems keeps increasing and eventually surpasses that of the former for voltages above approximately 1.5 V. This behaviour is quite unlike that of the fused MFBP junctions reported in Fig. 4, which have maximal currents that are significantly lower than the corresponding single MFBP junctions. Recent theoretical and experimental studies have demonstrated that edge-fused wires of metal-containing porphyrins and MFPs exhibit increasing bias-induced conductance with increasing length [38, 39]. This unintuitive phenomenon is due to the rapidly narrowing HOMO–LUMO gap as the length of the wire increases, countering the effect of longer tunneling distances. Our tentative results suggest that a similar effect could also occur in edge-fused MFBP molecular wires.

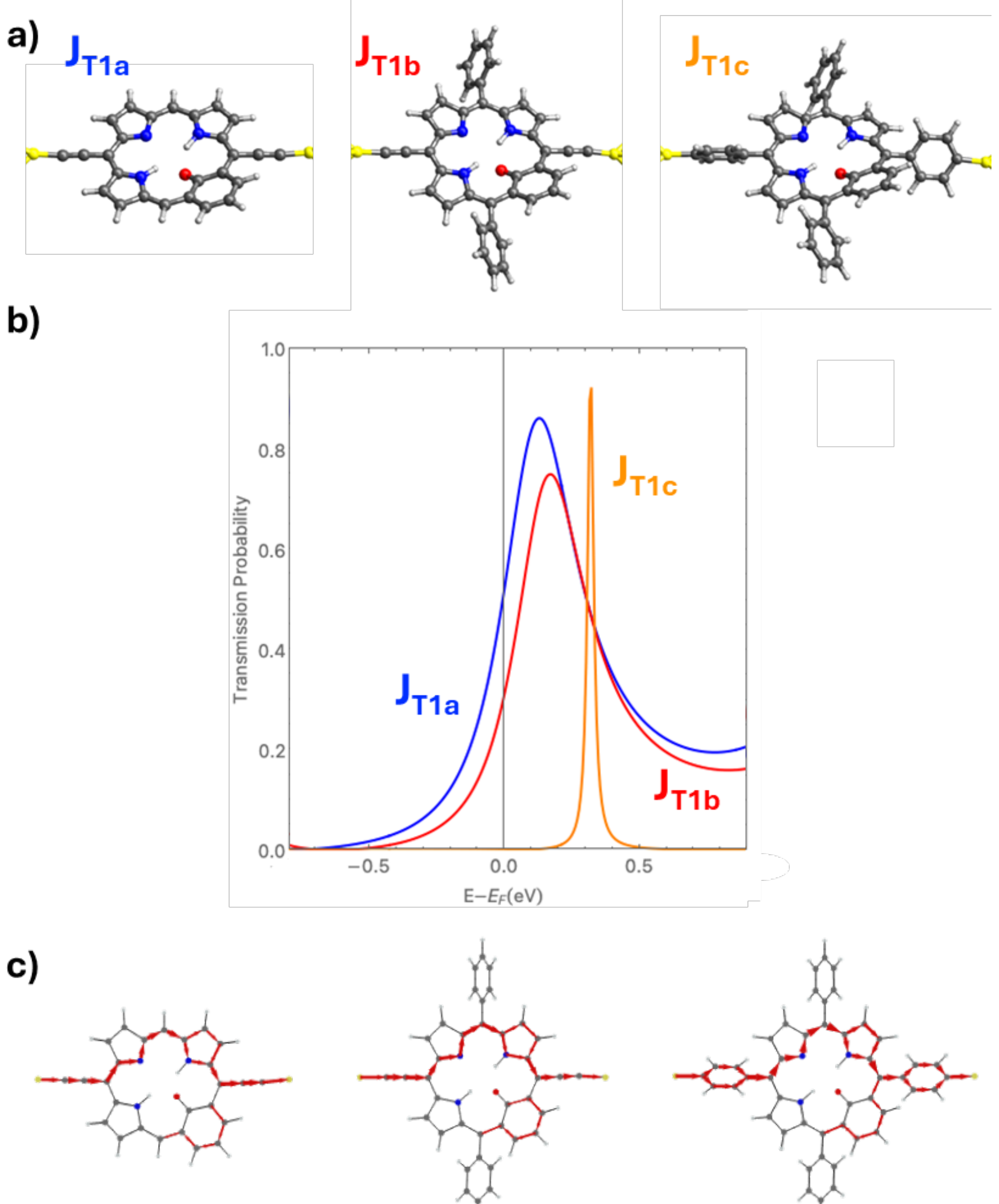

**Figure 5.** Comparison of the three variants of $\mathbf{J_{T1}}$. $\mathbf{J_{T1a}}$ - no aryl ring decoration and acetylenic electrode linkers, $\mathbf{J_{T1b}}$ - with aryl ring decoration and acetylenic electrode linkers, and $\mathbf{J_{T1c}}$ - aryl ring decoration and electrode linkers. Note that $\mathbf{J_{T1c}}$ is equivalent to $\mathbf{J_{T1}}$ in figure 3. For each case we show: a) structures of the MFBP with electrode anchors (atom colour key follows that in Fig. 3), b) transmission functions versus energy deviation from the Fermi level of the electrodes, c) local transport currents corresponding to the maximum transmission in b).

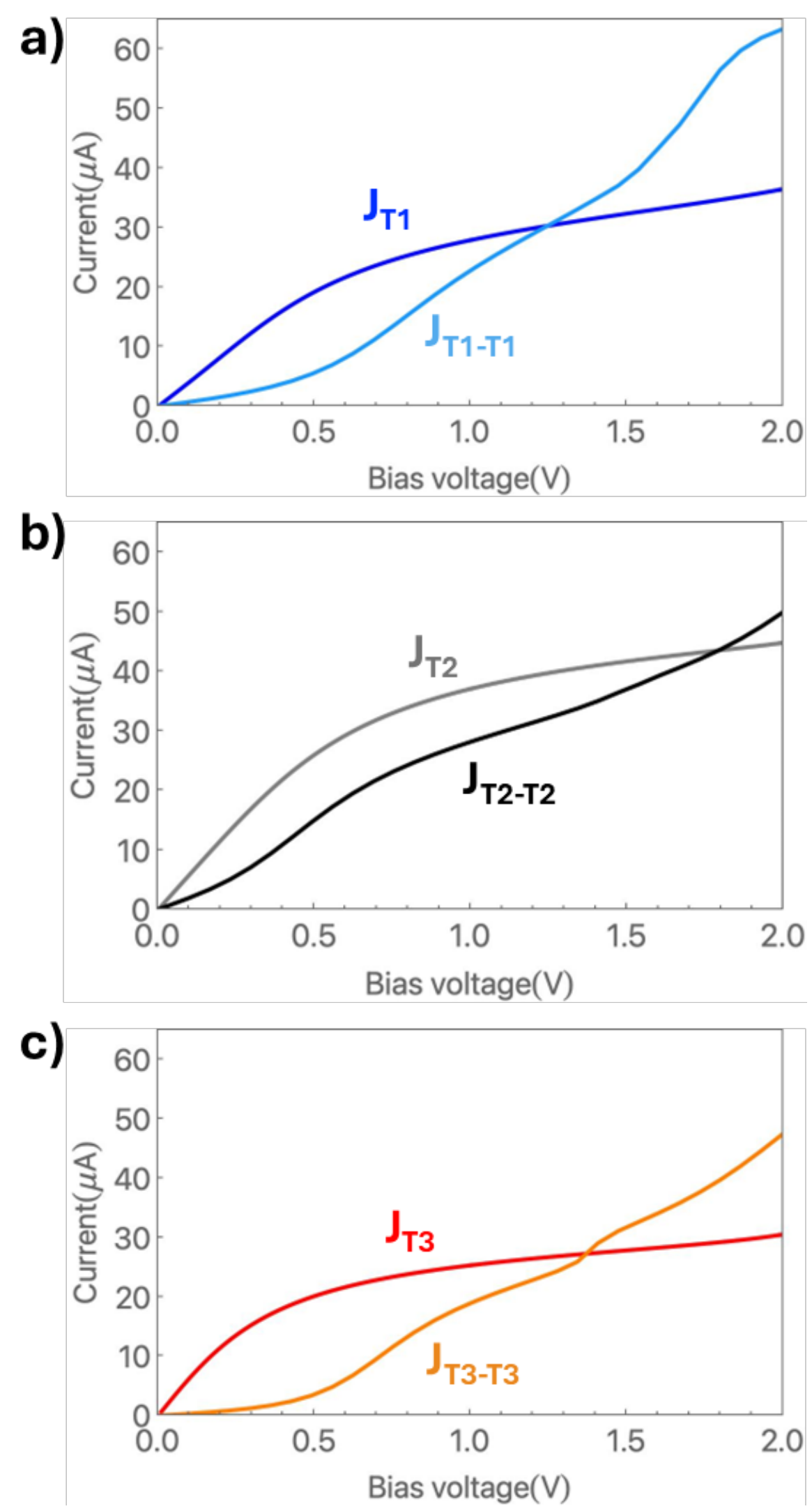

**Figure 6.** Comparison of current versus bias voltage characteristics for MFBP-based junctions with acetylenic electrode linkers: a) $\mathbf{J_{T1}}$ and $\mathbf{J_{T1\text{-}T1}}$, b) $\mathbf{J_{T2}}$ and $\mathbf{J_{T2\text{-}T2}}$ c) $\mathbf{J_{T2}}$ and $\mathbf{J_{T3\text{-}T3}}$.

## Conclusions

Our results demonstrate that E-fields can effectively bias tautomerization in asymmetric MFBPs. Experimentally achievable E-fields can selectively stabilize three distinct tautomers, each associated with unique aromatic or antiaromatic character and conductance signatures, providing a clear route for state detection. This direct coupling between tautomerism and aromaticity enables single MFBPs to function as three-state molecular switches with high ON/OFF ratios. Extending this concept, fused MFBPs offer multiple tautomeric configurations with distinct conductance profiles, creating opportunities for potential multistate molecular logic and memory elements. By tuning the electrode linkers, these systems can also exhibit wire-like transport with enhanced currents, suggesting a pathway toward MFBP-based molecular interconnects. Taken together, these findings position MFBPs as a versatile and scalable platform for E-field-responsive molecular electronics, bridging fundamental tautomeric control with practical device architecture potential.

## Acknowledgements

The authors acknowledge the financial support from the Spanish Ministerio de Ciencia, Innovación y Universidades and Agencia Estatal de Investigación (AEI) MCIN/AEI/10.13039/501100011033 through projects PID2023-149691NB-I00, PID2021-127957NB-I00, PID2024-157971NB-C22, TED2021-132550B-C21), Spanish Structures of Excellence María de Maeztu program (CEX2021-001202-M), and Agència de Gestió d'Ajuts Universitaris i de Recerca of Generalitat de Catalunya (2021SGR00354 and 2023CLIMA00064).